\documentclass[prl,twocolumn,superscriptaddress,footinbib]{revtex4-1} 

\usepackage{longtable}
\usepackage{morefloats}
\usepackage[dvips]{graphicx}
\usepackage{color}
\usepackage{epsfig,graphicx,amsfonts,amsbsy}
\usepackage{amsmath,amsfonts,amsthm,amssymb}
\usepackage{appendix}
\usepackage{makeidx}
\usepackage{url}
\usepackage{verbatim}
\usepackage[bookmarksnumbered,pdfpagelabels=true,plainpages=false,colorlinks=true,linkcolor=blue,citecolor=blue,urlcolor=blue]{hyperref}
\usepackage[rightcaption]{sidecap}
\usepackage{array}
\usepackage{xcolor,cancel}
\usepackage{multirow}
\usepackage{tabularx}
\usepackage{braket} 
\usepackage{dsfont}
\usepackage{ulem}
\usepackage{bbold}
\usepackage{etoolbox}

\newcommand{\pt}{\partial}
\newcommand{\mb}{\mathbf}
\newcommand{\mc}{\mathcal}

\newcommand{\Ed}{\mathcal{E}_{\scalebox{0.6}{D}}}

\begin{document}

\title{Curvature-Induced Skyrmion Mass}

\author{Alexander Pavlis}
\affiliation{ITCP and CCQCN, Department of Physics, University of Crete, 71003 Heraklion, Greece}
\affiliation{Institute of Electronic Structure $\&$ Laser, FORTH, 70013 Heraklion, Greece}
\author{Christina Psaroudaki}
\affiliation{Department of Physics, California Institute of Technology, Pasadena, CA 91125, USA}
\affiliation{Institute for Theoretical Physics, University of Cologne, D-50937 Cologne, Germany}

\date{\today}
\begin{abstract}
We investigate the propagation of magnetic skyrmions on elastically deformable geometries by employing imaginary time quantum field theory methods. We demonstrate that the Euclidean action of the problem carries information of the elements of the surface space metric, and develop a description of the skyrmion dynamics in terms of a set of collective coordinates. We reveal that novel curvature-driven effects emerge in geometries with non-constant curvature, which explicitly break the translational invariance of flat space. In particular, for a skyrmion stabilized by a curvilinear defect, an inertia term and a pinning potential are generated by the varying curvature, while both of these terms vanish in the flat-space limit.  
\end{abstract}

\maketitle
\textit{Introduction.--} The interplay between geometry, condensed matter order, and topology has been a rich source of novel physics throughout many disciplines, including thin magnetic materials \cite{Streubel_2016}, superfluid films \cite{PhysRevE.85.031150}, superconducting nanoshells \cite{PhysRevB.79.134516}, and nematic liquid crystals \cite{PhysRevLett.108.207803}. Recent advances in materials technology have made it possible to fabricate sub-micrometer sized systems with complicated geometry \cite{Tanda2002,PhysRevLett.84.2223,PhysRevLett.96.027205,Streubel2012}, and opened various possibilities toward tailoring physical phenomena at the nanoscale. In particular, considerable effort has been devoted to synthesizing magnetic nanostructures with modified curvature \cite{Sui2004,PhysRevLett.111.067202}, as they appear promising elements in high-density magnetic memories \cite{Hrkac2011,fernandezpacheco2017}. 

The relation between topological defects and curved surfaces can be a source of new geometric effects \cite{Bausch1716,Nelson2002,PhysRevLett.93.215301,Irvine2010}. Specifically to magnetism, the coupling between surface order and curvature plays an important role in the stability and dynamical properties of magnetization textures \cite{PhysRevLett.114.197204,PhysRevLett.112.257203,Sheka_2015,Sheka2020}, and the band structure of magnon modes \cite{Gaididei2018,Korniienko2019}. This coupling is particularly evident in two-dimensional systems known to host topological excitations \cite{zang_topology_2018,RevModPhys.82.1301}. Among such excitations, magnetic skyrmions, particle-like topological textures, have attracted much attention due to potential applications in magnetic information storage and processing devices \cite{Nagaosa2013,Wiesendanger2016}. Although skyrmion stability has been extensively studied for a variety of curvilinear surfaces, including cylindrically symmetric curved surfaces \cite{CARVALHOSANTOS20131308,CARVALHOSANTOS2015179}, spherical shells \cite{PhysRevB.94.144402}, curvilinear defects \cite{PhysRevLett.120.067201}, and curvature gradients \cite{PhysRevApplied.10.064057}, their dynamics has not been addressed before. Similarly to the dynamics of domain walls \cite{Landeros2010,PhysRevB.92.104412,PhysRevB.98.060409}, it is expected that local changes in the geometry of the surface will result in remarkable changes in the dynamic properties. 

The present paper aims to develop a formalism to describe the dynamics of a skyrmion propagating on a magnetic surface with nontrivial geometry. Skyrmions emerge as topological solutions of the magnetization field, usually parametrized by a number of collective coordinates of position. Here, we introduce a novel coupling of the skyrmion with the underlying curvature and demonstrate that spaces with non-constant curvature generate an effective potential and an inertial term for the skyrmion guiding center. In particular, we explicitly calculate a position-dependent mass term for a skyrmion stabilized by a curvilinear defect, which scales with the skyrmion radius. Both the mass and the potential vanish in the flat-space limit. 

\textit{Field Theory in Curved space.--} We consider an arbitrary curvilinear ferromagnetic insulating shell, with normalized magnetization $\mb{m}$ valued on a unit sphere in cartesian coordinates, $\mb{m} =  \sin \Theta \cos \Phi \mb{x}+ \sin \Theta \sin \Phi \mb{y}+\cos \Theta \mb{z}$, described by the imaginary time Euclidean action  
\begin{align}
\mc{S}_E =\frac{L}{\alpha} \int_{0}^{\beta} d\tau \int d\mc{A}~ [\frac{i S l^2}{\alpha^2} \dot{\Phi}(1-\Pi) +\mc{W}(\Phi,\Pi) ]\,.
\label{eq:EucAction}
\end{align}
Here $S$ is the magnitude of the spin, $\alpha$ is the lattice spacing, $\Pi=\cos \Theta$, $L$ is the film thickness, and the dot denotes a time derivative, $\dot{\Phi} = \pt_\tau \Phi$, a convention we adopt from now on. Also, we set $\hbar=1$ throughout. The surface is parametrized by the curvilinear coordinates $(\eta_1,\eta_2$), given here in dimensionless units. We introduce $l$ as a model dependent magnetic length and $d\mc{A} = \sqrt{\vert g \vert} d\eta_1 d\eta_2$ the surface element with $\sqrt{\vert g \vert} = \sqrt{\vert \det[g_{ab}]\vert}$ and $g_{ab}$ the surface space metric. The corresponding curvilinear vector components are denoted as $\mb{e}_{1}$ and $\mb{e}_2$, with $\mb{e}_a \cdot \mb{e}_b = \delta_{ab}$. The partition function is given by a functional integral, $Z = \int \mc{D} \Phi \mc{D} \Pi e^{-\mc{S}_E}$. The magnetization field can be decomposed in the local orthonormal basis as $\mb{m} =  \sin \Theta_c \cos \Phi_c \mb{e}_1+ \sin \Theta_c \sin \Phi_c \mb{e}_2 +\cos \Theta_c \mb{n}$, where $\mb{n} = \mb{e}_1 \times \mb{e}_2$ is the unit vector normal to the surface. We note, however, that the dynamical part of the action \eqref{eq:EucAction} assumes that the fields $\Phi$ and $\Pi$ are magnetization components in cartesian coordinates, a convenient basis to employ the transformation properties of the SU(2) coherent-state representation used to derive the path integral for the spin algebra \cite{PhysRevD.19.2349,Kochetov1995, PhysRevB.53.3237}. 

Our current task is to describe the dynamics of topologically nontrivial magnetization textures, stabilized by the energy functional $\mc{W}$, which for now we keep general. Magnetic skyrmions are characterized by a finite topological charge $Q=(1/4\pi) \int q (\mb{m})~d\mc{A}$, with the topological density
\begin{align}
q(\mb{m})= -\frac{\epsilon_{ab}}{2}\mb{m} \cdot [(\nabla_a \mb{m}) \times (\nabla_b \mb{m})] \,. 
\end{align}
The index $Q$ describes the degree of the map from an arbitrary curvilinear surface into a sphere $S^2$ \cite{PhysRevB.94.144402,PhysRevLett.51.2250}. Here $\epsilon_{ab}$ is the Levi-Civita tensor, $\nabla_a = (g_{aa})^{-1/2} \pt_a$, and summation over repeated indices is implied. Here we consider a class of surfaces in which the magnetization field obeys the relevant homotopy group $\Pi_2(S^2)=\mathbb{Z}$ which ensures an integer topological charge \cite{PhysRevB.77.134450}, and surfaces with a topological excitation characterized by a vanishing density $q(\mb{m})$ away from the skyrmion core. $Q$ is conserved and an integer-valued for any closed surface $\mc{A}$ (see Ref.\cite{PhysRevB.94.144402}), as it corresponds to the difference of negatively and positively charged monopoles inside that surface \cite{1979iv}. Here we generalize earlier considerations derived for vortices on planar films \cite{PAPANICOLAOU1991425} to demonstrate that the topological charge of a skyrmion on an arbitrary curvilinear shell is preserved by the dynamics. The Euler Lagrange equation of Eq.~\eqref{eq:EucAction} takes the form $\dot{\mb{m}} + \mb{m} \times (\delta E /\delta \mb{m})=0$, obtained upon replacing imaginary time $\tau$ with real time $t=-i \tau$, and $E=(L/\alpha) \int \mc{W}(\mb{m})~d\mc{A} $. By using vector calculus identities, the time evolution of the topological charge may be written in the form of a local conservation law, $\dot{q} = \epsilon_{ab} \nabla_a (\mb{F} \cdot \nabla_b \mb{m})$, resulting the global conservation law $\dot{Q}=0$, for any choice of the effective magnetic field $\mb{F} = \delta E/\delta \mb{m}$ and the metric tensor $g_{ab}$. The topological charge conservation suggests that skyrmions maintain their soliton-like character under rigid translations, in contrast to Bose-Einstein condensates \cite{Khaykovich1290}, where the standard notion of soliton is lost in the presence of translational symmetry breaking terms and is only maintained in homogeneous and isotropic spaces with a constant curvature \cite{PhysRevA.81.053806}.

\begin{figure}[t!]
\includegraphics[width=1\linewidth]{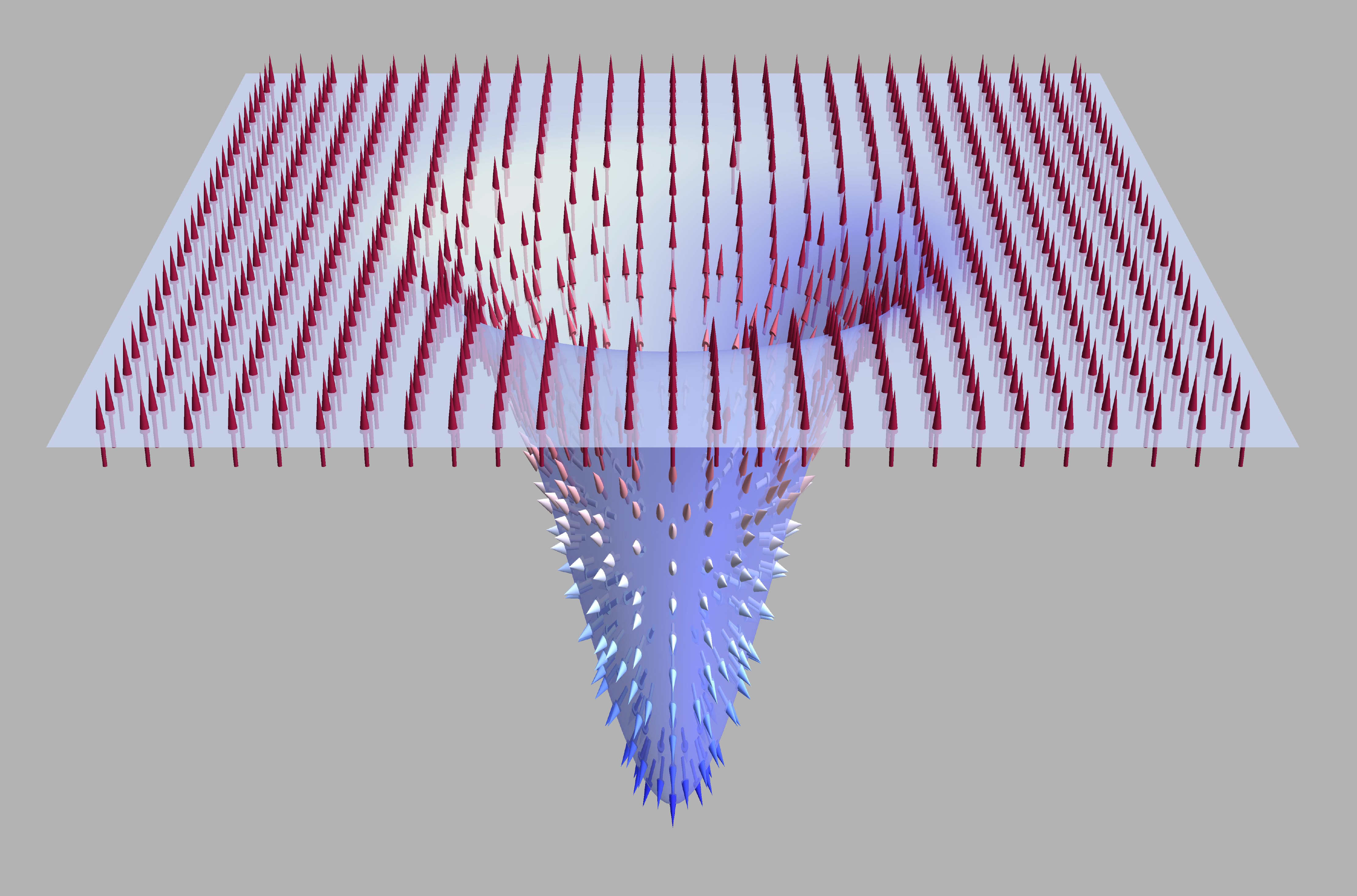} 
\caption{Skyrmion profile with a topological charge $Q=1$, realized in a curvilinear concave Gaussian defect, with the skyrmion center located at the defect center.}
\label{fig:defect} 
\end{figure}

\begin{figure*}[t] 
\includegraphics[width=1\textwidth]{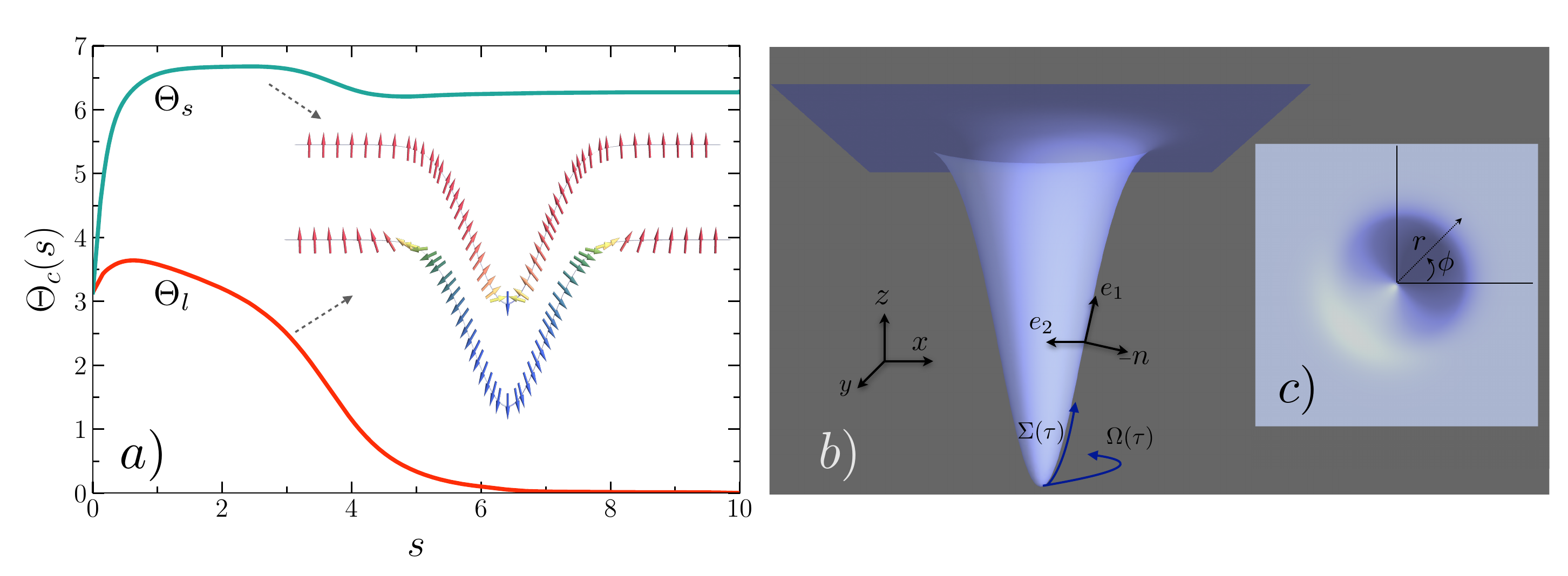}
\caption{(a) Skyrmion profile $\Theta_c(s)$ for two skyrmion radii, a large $\Theta_l$ and a small one $\Theta_s$, with chosen $U_0=-3$, $r_0=1$, and $d=1.1$, while a vertical cross section view of the magnetization profile is depicted for both skyrmion sizes. b) The geometry of the Gaussian defect determined by the function $z(r)=U_0e^{-r^2/(2 r_0)^2}$. The curvilinear basis is indicated by the vectors $\{\mb{e}_1, \mb{e}_2,\mb{n} \}$, while $\Sigma(\tau)$ denotes the collective coordinate in the arc direction, and $\Omega(\tau)$ in the azimuthal direction. c) Upper view of the curvilinear surface, where $r$ denotes the radial coordinate, and $\phi$ is the azimuthal angle of rotations.}
\label{fig:Merge}
\end{figure*}   

We now promote the collective coordinates of position to dynamical variables by considering the spin field of a moving skyrmion as $\mb{m}(\mb{r},\tau) = \mb{m}_0[\mb{r}-\mb{R}(\tau)]$, where $\mb{m}_0$ is a static solution of $\delta \mc{S}_E =0$ for $\mb{R}=0$. By inserting this ansatz into the action $S_E$ of Eq.~\eqref{eq:EucAction} and considering small perturbations in the path $\mb{R}$, we arrive at
\begin{align}
\mc{S}^{0}_E = \int_0^{\beta} d\tau [-i\tilde{Q} (R_1 \dot{R}_2 - R_2 \dot{R}_1) + V(\mb{R})] \,,
\label{eq:ActionColl}
\end{align}
with $\tilde{Q}=2 \pi S L l^2Q/\alpha^3$. The first term corresponds to the well known Magnus force \cite{PhysRevB.53.16573,PhysRevLett.30.230} proportional to the skyrmion velocity, while the geometric potential $V$ originates from the energy term $\mc{W}$. Next, one can consider fluctuations around the skyrmion configuration as $\mb{m}(\mb{r},\tau)= \mb{m}_0[\mb{r}-\mb{R}(\tau)] + \delta \mb{m}[\mb{r}-\mb{R}(\tau),\tau]$, and perform finite perturbation theory in terms of $\delta \mb{m}$. In the flat-space limit, the interaction of the skyrmion with the surrounding magnon modes gives rise to a mass term when potentials that break translational symmetry arising from defects, nonuniform magnetic fields, and spatial confinement \cite{PhysRevX.7.041045,PhysRevLett.120.237203,doi:10.1021/nl501379k,PhysRevApplied.12.064033} are present. In these studies, the external potential is treated perturbatively, and the resulting mass is proportional to $U_0^2$, with $U_0$ the potential strength. The presence of a mass term for skyrmions subjected to magnetic circular disks has been experimentally reported in Ref.~\cite{Buttner2015}. Below we demonstrate, without employing perturbative methods, that in a curvilinear surface with non-constant curvature, such as a curvilinear defect depicted in Fig.~\ref{fig:defect}, the skyrmion-magnon interaction gives rise to a position-dependent mass, while both the mass and the pinning potential vanish in the flat-space limit. 

\textit{Curvilinear Defect.--} In the following, we consider an energy of the form $E=(L/\alpha) \int \mc{W} d\mc{A}$, with
\begin{equation}
\mc{W}= J [\nabla_a \mb{m} \cdot \nabla_a \mb{m} + 1-(\mb{m}\cdot \mb{n})^2 + d \Ed] \,,
\label{eq:EnFunc}	
\end{equation}
realized in a curvilinear defect originally introduced in Ref.~\onlinecite{PhysRevLett.120.067201}, with details repeated here for completeness. The magnetic length of the model is $l=\sqrt{J/K}$, where $J$ in units of energy denotes the exchange coupling and $K$ is the anisotropy coupling. We introduce the dimensionless $d=D/\sqrt{JK}$ as the coupling of the Dzyaloshinskii-Moriya interaction, $\Ed =(\mb{m}\cdot \mb{n}) \nabla \cdot \mb{m} - \mb{m}\cdot \nabla (\mb{m}\cdot \mb{n})$, originating from inversion-symmetry breaking \cite{PhysRevB.94.144402,PhysRevLett.112.257203,Sheka_2015}. The Gaussian defect is determined by $z(r) = U_0 e^{-r^2/2 r_0^2}$, with $r$ the radial coordinate, $U_0$ the amplitude, and $r_0$ the width of the defect (see Fig.~\ref{fig:Merge}-(b) for details of the considered geometry). The two principal curvatures $k_1=g(r)^3 z''(r)$ and $k_2=g(r) z'(r)/r$ determine the properties of the surface, where $g(r)=1/\sqrt{1+z'(r)^2}$. The local orthonormal basis $\{ \mb{e}_1,\mb{e}_2 ,\mb{n} \}$ is defined by unit vectors expressed in the cartesian basis as $\mb{e}_1= g(r) \{ \cos \phi,\sin \phi,z'(r) \}$, $\mb{e}_2 = \{ -\sin \phi, \cos \phi,0 \}$, and $\mb{n}= g(r)\{-z'(r) \cos \phi,-z'(r) \sin \phi, 1\}$.  Here $r$ is the radial and $\phi$ the azimuthal coordinate (see Fig.~\ref{fig:Merge}-(c)). Finally, instead of $r$, the field configuration is expressed in terms of a coordinate $s$ along the arc of the Gaussian, and $r=r(s)$ is determined by the set of equations $r'(s)^2(1+z'[r(s)]^2])=1$ and $r(0)=0$. Figs.~\ref{fig:Merge}-(b)-(c) summarize the properties of the curvilinear geometry. 

In terms of the parametrization $\mb{m} =  \sin \Theta_c \cos \Phi_c \mb{e}_1+ \sin \Theta_c \sin \Phi_c \mb{e}_2 +\cos \Theta_c \mb{n}$, the static stable solutions determined by $\delta E/\delta \Theta_c =0$ and $\delta E/\delta \Phi_c =0$, correspond to $\Phi_c=0,\pi$ and the rotationally symmetric solution $\Theta_c = \Theta_c (s)$ with boundary conditions $\Theta_c(0) = 0$ and $\Theta_c(\infty) = 0,2\pi$, satisfying the equation,
\begin{align}
\nabla_s^2\Theta_c -\sin \Theta_c \Xi/2 +\mc{C}r'(d-2k_1)\sin^2 \Theta_c /r =\mc{C}(k_1'+k_2') \,.
\label{eq:EqTheta}
\end{align}
Here $\mc{C}=\cos \Phi_c=\pm1$, $\nabla_s^2\Theta_c = (r \Theta_c')'/r $, and $\Xi=1+r'^2/r^2 -k_2^2+d (k_1+k_2)$. Magnetization profiles of skyrmions, obtained by a numerical solution of Eq.~\eqref{eq:EqTheta}, are depicted in Fig.~\ref{fig:Merge}, for $U_0 =-3$ and $r_0=1$, and $d=1.1$ \cite{PhysRevLett.120.067201}. For this choice of parameters, both of these solutions are characterized by $Q=g(r)\cos \Theta_c/2 \vert_0^{\infty} =1$ and no displacement instability is present. The large radius skyrmion $\Theta_l$ (red line), is the state with the lowest energy $E=-9.1 J L/\alpha$, the small radius skyrmion $\Theta_l$ (green line) has $E=32.3 JL/\alpha$, and the uniform state has $E=7.5JL/\alpha$ \cite{PhysRevLett.120.067201}. This is in contrast to the planar skyrmion with $Q=1$, which always appears as an excitation above the ferromagnetic background. 

For the specific geometry considered here, we find that the fields in cartesian coordinates are related to the ones in curvilinear as $\Phi = \phi$ (for $\Phi_c=0$) and $\Pi= g(r) (\Pi_c - z'(r)\sqrt{1-\Pi_c^2})$. We now consider deviations of the field $\Phi_c =\Phi_c^0+\xi$, with $\vert \xi \vert \ll1$ and $\Phi_c^0 =0$ for the remainder of the paper. The action \eqref{eq:EucAction}, up to second order in $\xi$, is
\begin{align}
\mc{S}_E= S_E^0+ (L/\alpha) \int_0^{\beta} d\tau \int d\mc{A} ( \mc{J}\xi + \xi \mc{L}_{\xi} \xi)\,,
\end{align}
where $S_E^0 = L \int_{\tau,\mc{A}} [(iSl^2/\alpha^3)\dot{\Phi}_0(1-\Pi)+\mc{W}_0(\Pi_c)]$, with $\Phi_0=\phi$, and we assume $\beta$-periodic in time fluctuating fields, $\xi(0)=\xi(\beta)$. Here we introduce $\mc{J} = (iSl^2/\alpha^2) \dot{\Pi}_c$. 
In the planar film limit $g(r)=1$, $z'(r)=0$, and $\Pi=\Pi_c$. The detailed forms of the energy functional $\mc{W}_0$ and the operator $\mc{L}_\xi$ are given in Ref.~\cite{SuppNote}. It is worth mentioning that once we introduce collective coordinates of position as $\Phi_0(\mb{r},\tau)= \Phi_0[\mb{r}-\mb{R}(\tau)]$ and $\Pi_0(\mb{r},\tau)= \Pi_0[\mb{r}-\mb{R}(\tau)]$, $\mc{S}^0_E$ is given by \eqref{eq:ActionColl}, which can be written in the equivalent form $\mc{S}^0_E = \int_\tau[ -i \tilde{Q}\Sigma(\tau)^2  \dot{\Omega}(\tau)+V(\Sigma)]$, with $\Sigma$ the collective coordinate along the arc of the defect, $\Omega$ in the azimuthal direction (see Fig.~\ref{fig:Merge}-(b)), and $V(\Sigma)$ presented below. We focus on the dynamics of a skyrmion located at the center of the defect, with constrained dynamics $\dot{\Omega} \simeq 0$ and $\mc{S}_E^0 \simeq L \int_{\tau, \mc{A}} \mc{W}_0(\Pi_c[s-\Sigma(\tau) ])\simeq \int_\tau V(\Sigma)$. 
\begin{figure}[t]
\centering
\includegraphics[width=1\linewidth]{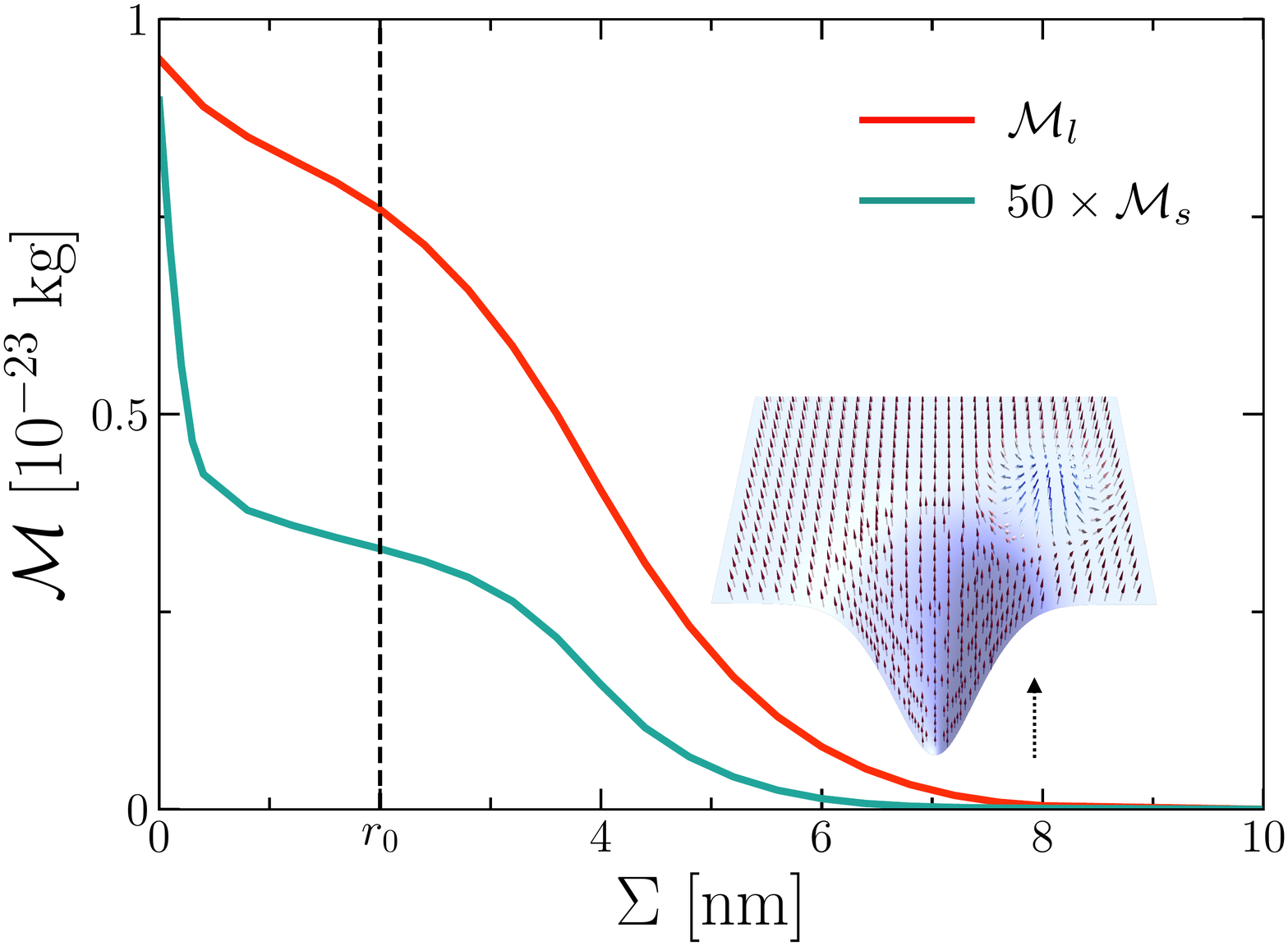}
  \caption{Curvature induced mass $\mc{M}$ as a function of the collective coordinate $\Sigma$ in physical units. $\Sigma$ represents the distance between the center of the defect and the center of the skyrmion. $\mc{M}_l$ ($\mc{M}_s$) denotes the mass for the large (small) radius skyrmion, and the vertical dashed line indicates the defect width $r_0$. }
\label{fig:Mass} 
\end{figure}

 We integrate out the $\xi$ fluctuations from the partition function $Z$ by noting that the path integral measure is replaced by $\mc{D} \Phi \rightarrow \mc{D} \xi$, and that the integral can be written in a Gaussian form by shifting the fluctuating fields by $\tilde{\xi} = \xi +1/2 \mc{J} \mc{L}^{-1}_\xi$. The partition function takes the form, $Z = \int \mc{D} \Pi [e^{-\mc{S}}/\det[(L/\alpha)\mc{L}_{\xi}]$, with the new action given by $\mc{S}= \mc{S}_E^0 -(1/4)\mc{J} \circ\mc{L}^{-1}_{\xi} \mc{J}$, where we introduce the compact notation $\mc{J} \circ \mc{L}^{-1}_{\xi} \mc{J} = (L/\alpha)\int_{0}^{\beta} d\tau \int d\mc{S} d\mc{S'} \mc{J}(\mb{r},\tau) \mc{L}^{-1}_{\xi} (\mb{r},\mb{r}')\mc{J} (\mb{r}',\tau)$. The operator $\mc{L}^{-1}_{\xi}$ can be expanded in the space of the eigenfunctions $\Psi_n$, solutions of the eigenvalue problem $\mc{L}_{\xi} \Psi_{n} = \epsilon_n \Psi_{n}$. The dynamical part of the action contains terms $\mc{O}(\dot{\Pi}^2)$, which correspond to inertia terms for the collective coordinate of translations in the arc direction, $\Pi_{c}=\Pi_c[s-\Sigma(\tau)]$. Since the model is not translational invariant, $\Sigma(\tau)$ is energy dependent and the eigenenergies of $\mc{L}_\xi$ are positive $\epsilon_n >0$ \cite{PhysRevLett.120.067201}. We then find that 
 \begin{align}
 \mc{S}_E = \int_0^\beta d\tau [\frac{1}{2} \mc{M}(\Sigma) \dot{\Sigma}^2 + V(\Sigma) ]\,,
\end{align}   
with a position dependent mass term given by $\mc{M}(\Sigma)  = (S^2 L l^4/2 \alpha^5) \sum_n \vert M_n(\Sigma)\vert ^2/\epsilon_n$, and matrix elements $M_n(\Sigma) = \int \pt_s \Pi_c[s-\Sigma] \Psi_n(\mb{r})d\mc{A}$. As expected, due to the lack of translation symmetry, the effective mass depends on the background geometry and thus on the collective coordinate $\Sigma$. We assume that the fluctuations have a larger wavelength than the skyrmion radius, such that the geometric potentials arising from the underlying varying curvature become the dominant terms in the inertia integral. In this limit $\mc{L}_\xi^{-1} \simeq [J U(s)]^{-1}$, with $U(s) = \Theta_c' k_1 + \sin \Theta_c \cos \Theta_c k_2 r'/r+ (\cos^2 \Theta_c  -1)(k_1^2-k_2^2)$. We then arrive at the simplified form
\begin{align}
\mc{M}(\Sigma) \simeq \frac{S^2 L l^4}{2J \alpha^5} \int \frac{[\pt_s \Pi_c(s-\Sigma)]^2}{U(s)} d\mc{A} \,.
\label{eq:Mass}
\end{align}

The dependence of $\mc{M}$ on the collective coordinate $\Sigma$, which represents the distance between the center of the defect and the center of the skyrmion, is summarized in Fig.~\ref{fig:Mass}. Both $\mc{M}$ and $\Sigma$ are given in physical units, with $L=3 \alpha$, $l=4\alpha$, $\alpha= 5$ \AA, and $J=2$ meV. As expected, we find that $\mc{M}$ decreases as the skyrmion departs from the defect center, suggesting that the mass vanishes once translation symmetry is restored. It is also apparent that $\mc{M}$ grows with the skyrmion size, as $\mc{M}_l \gg \mc{M}_s$, where $\mc{M}_{l,s}$ is the mass calculated for the large (small) radius skyrmion respectively. The role of fluctuations around the field $\Pi$ will give rise to mass renormalization terms that are descibed in detail in Ref.~~\cite{SuppNote}. 
\begin{figure}[t]
\centering
\includegraphics[width=0.98\linewidth]{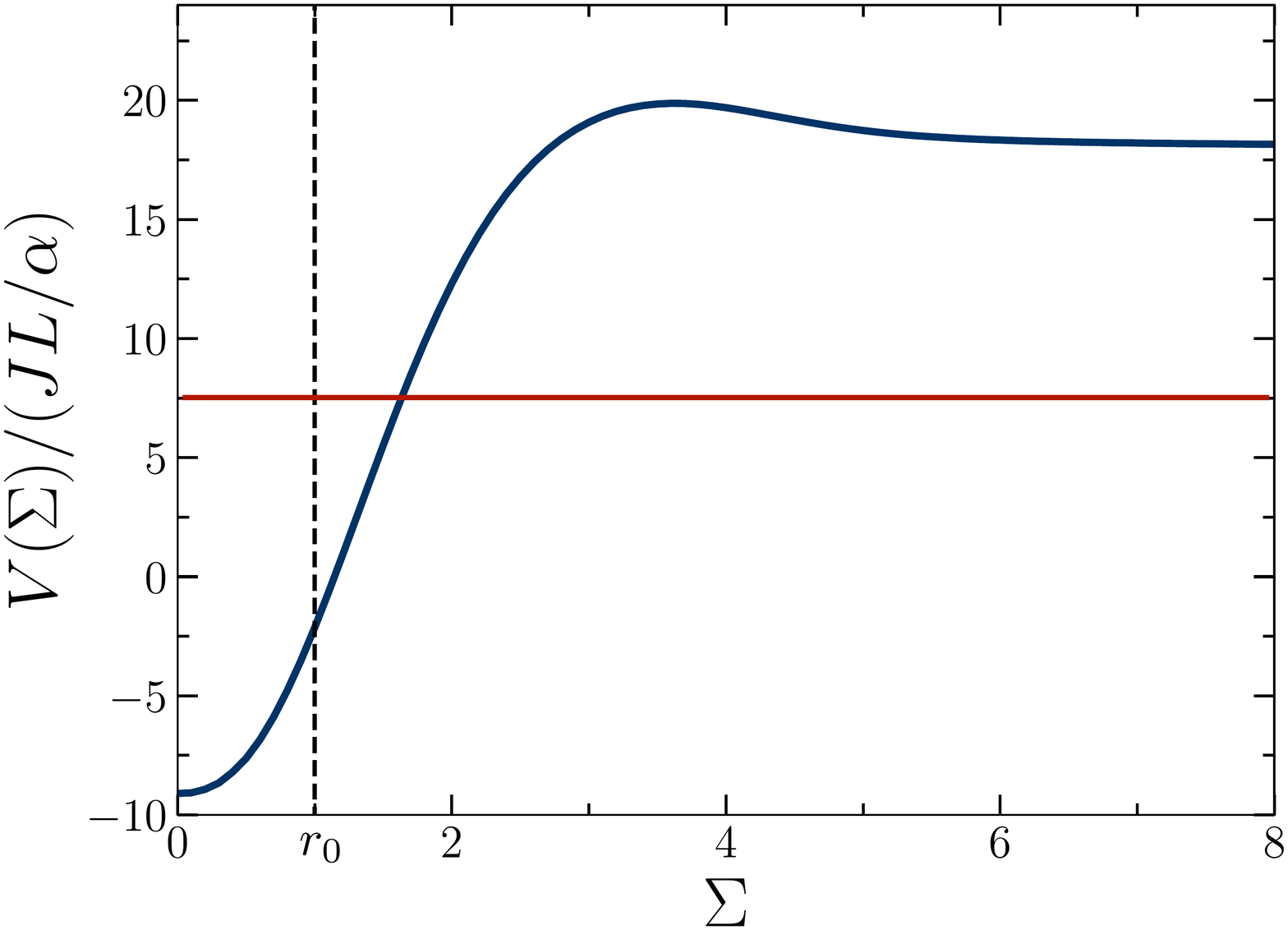}
  \caption{Pinning potential $V$ as a function of the collective coordinate $\Sigma$, representing the distance between the center of the defect and the center of the skyrmion, in dimensionless units. The vertical dashed line indicates the defect width $r_0$, while the horizontal red line indicates the energy of the ferromagnetic background.}
\label{fig:PinPot} 
\end{figure}

To complete the description, we must also examine the pinning potential, defined as 
\begin{align}
V(\Sigma) = \frac{L}{\alpha} \int \mc{W}_0 [\Pi_c(s-\Sigma)] d\mc{A} \,,
\end{align}
where explicit formulas for $\mc{W}_0$ are provided in Ref.~\cite{SuppNote}. The potential is depicted in Fig.~\ref{fig:PinPot}, calculated for the large radius skyrmion $\Theta_l$, as a function of the distance between the defect and skyrmion center. It is apparent that a local bend on the surface is a source of pinning, in analogy to curvature-induced pinning potentials already predicted for domain walls in magnetic nanowires \cite{PhysRevB.92.104412,doi:10.1063/1.3246154,doi:10.1063/1.3062828}. The energy of the planar skyrmion for $\Sigma \gg 1$ is larger than of the ferromagnetic state (red solid line), suggesting that the skyrmion with $Q=1$ is no longer the ground state, but appears as an excitation above the uniform background. 

We should emphasize that, although the mass $\mc{M}$ vanishes when the skyrmion is displaced away from the defect (thus when the translational symmetry is restored locally) the value of $\mc{M}$ predicted from \eqref{eq:Mass} diverges when $k_1,k_2 \rightarrow 0$. In this limit, global translational symmetry is recovered, and the collective coordinates represent the zero-energy modes associated with translations. To properly quantize the skyrmion system, one needs to, not only elevate $\mb{R}$ to a dynamical variable, but also introduce gauge fixing constraints in the path integral, to remove the singularities that originate from overcounting degrees of freedom \cite{PhysRevD.11.2943,PhysRevD.12.1038,doi:10.1142/0163,Rajaraman:1982is}. Thus, divergences are treated by imposing the so-called rigid gauge, a constraint that requires that the zero modes are orthogonal to the fluctuating fields. For our purposes, it suffices to note that the theory constructed here assumes a finite curvature, while the skyrmion propagation in a 2D planar film has been treated elsewhere \cite{PhysRevX.7.041045}. Our present investigation suggests that curvature in thin magnetic fields introduces new ways to tailor, not only the static but the dynamic properties of magnetic topological particles as well, an effect that we anticipate to be of high importance for nanomagnetism applications. 
 
\begin{acknowledgments} A.P. acknowledges helpful discussions with T.N. Tomaras. A.P. was supported by the Onassis foundation and the Institute for Theoretical and Computational Physics - ITCP (Crete). C.P. has received funding from the European Union's Horizon 2020 research and innovation programme under the Marie Sklodowska-Curie grant agreement No 839004. 
\end{acknowledgments}

\bibliography{Skyrmion_Curvature}
 
\end{document}